\documentclass[a4paper]{article}
\usepackage{xcolor}
\usepackage[utf8]{inputenc}
\usepackage[T1]{polski}
\usepackage{graphicx}\graphicspath{{img/}}
\usepackage{lmodern,fancyhdr,secdot,float,amsmath,amsfonts,amsthm,amssymb}
\usepackage{booktabs,tabularx}
\usepackage[margin=2cm]{geometry}
\usepackage{multicol}
\usepackage[hidelinks,unicode]{hyperref}
\usepackage{titlesec}
\titleformat{\section}{\centering\normalsize\bf}{\thesection.}{.5em}{\MakeUppercase}
\titleformat*{\subsection}{\bf\normalsize\selectfont}
\titleformat*{\subsubsection}{\bf\normalsize\selectfont}
\newcommand{\titlePL}[1]{\large\textbf{ #1}}
\newcommand{\titleEN}[1]{\normalsize #1}
\newcommand{\keywordsPL}[1]{\small\textbf{Słowa kluczowe:} #1}
\newcommand{\keywordsEN}[1]{\small\textbf{Keywords:} #1}
\newcommand{\abstractPL}[1]{\small\textbf{Streszczenie:} #1}
\newcommand{\abstractEN}[1]{\small\textbf{Abstract:} #1}

\definecolor{logo_color}{RGB}{40, 69, 166}

\begin{document}\thispagestyle{empty}\pagestyle{fancy}
\begin{minipage}[t]{0.5\textwidth}\vspace{0pt}%
\includegraphics[scale=0.9]{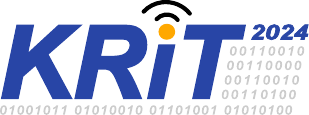}
\end{minipage}
\begin{minipage}[t]{0.45\textwidth}\vspace{12pt}%
\centering
\color{logo_color} KONFERENCJA RADIOKOMUNIKACJI\\ I TELEINFORMATYKI\\ KRiT 2024
\end{minipage}

\vspace{1cm}

\begin{center}
\titlePL{ROZPROSZONE WYKRYWANIE ZAJĘTOŚCI WIDMA OPARTE NA UCZENIU FEDERACYJNYM}

\titleEN{DISTRIBUTED SPECTRUM OCCUPANCY DETECTION BASED ON FEDERATED LEARNING}\medskip

Łukasz Kułacz$^{1}$, 
Adrian Kliks$^{2}$

\medskip

\begin{minipage}[t]{0.6\textwidth}
\small $^{1}$ Politechnika Poznańska, Poznań, \href{mailto:email}{lukasz.kulacz@put.poznan.pl} \\
\small $^{2}$ Politechnika Poznańska, Poznań, \href{mailto:email}{adrian.kliks@put.poznan.pl} 
\end{minipage}

\medskip

\end{center}

\medskip

\begin{multicols}{2}
\noindent
\abstractPL{Wykrywanie zajętości widma jest kluczowym zagadnieniem umożliwiającym dynamiczny dostęp do widma. Współcześnie w celu polepszenia detekcji popularne są rozwiązania z obszaru uczenia maszynowego, w tym uczenia federacyjnego (FL). Głównym wyzwaniem w tym kontekście jest ograniczony dostęp do danych treningowych. W pracy przedstawiono podejście rozproszone FL, skupiając się na węzłach pozbawionych dostępu do danych uczących. Omówiono wyniki eksperymentu sprzętowego polegającego na wykrywaniu sygnału DVB-T.}
\medskip

\noindent
\abstractEN{
Spectrum occupancy detection is a key enabler for dynamic spectrum access, where machine learning algorithms are successfully utilized for detection improvement. However, the main challenge is limited access to labeled data about users' transmission presence needed in supervised learning models. We present a distributed federated learning approach that addresses this challenge for sensors without access to learning data. The paper discusses the results of the conducted hardware experiment, where FL has been applied for DVB-T signal detection. }
\medskip

\noindent
\keywordsPL{detekcja zajętości widma, dynamiczny dostęp do widma, uczenie federacyjne, uczenie maszynowe}
\medskip

\noindent
\keywordsEN{spectrum occupancy detection, dynamic spectrum access, federated learning, machine learning}

\section{Wstęp}

Statycznie i przydzielone na wyłączność widmo gwarantuje proste zarządzanie zasobami częstotliwości, ale jednocześnie jest dalekie od optymalnego. Liczne prace dowiodły, że takie podejście może prowadzić do tylko częściowego wykorzystania dostępnych zasobów widmowych i nie rozwiązuje problemu nieefektywnego wykorzystania widma. Dynamiczny dostęp do widma (ang. Dynamic Spectrum Access, DSA) jest jednym z obiecujących rozwiązań tego problemu \cite{Arjoune2019, Liang2011, dudda2012}. Wykorzystując właściwość elastyczności w dostępie do zasobów widmowych, DSA pozwala on na wykorzystanie niezajętego widma do celów innych niż te, dla których zostało ono pierwotnie przeznaczone i~zarezerwowane.\\ 
Zagadnienie poprawnej identyfikacji niezajętych fragmentów widma częstotliwościowego, choć koncepcyjnie proste, jest trudne w realizacji.  Wiele rozwiązań zostało zaproponowanych w dziedzinie wykrywania widma (ang. Spectrum Sensing, SS) \cite{Haykin2005,Yucek2009}, gdzie ostatnio rola sztucznej inteligencji (ang. Artificial Intelligence, AI) staje się coraz bardziej dominująca~\cite{elec2021}. Jednak w~przypadku zastosowania uczenia maszynowego, istotnym wyzwaniem staje się sam proces uczenia, trenowania i weryfikacji modeli AI. Jednym z możliwych rozwiązań tego problemu przekazywanie pomiędzy węzłami w systemie SS informacji o~parametrach modelu. Rozwiązanie takie jest możliwe np. w~uczeniu federacyjnym (ang. Federated Learning, FL) czy tzw. transfer learning~\cite{sensors2022, iot2021}.\\
W \cite{kulaczSensors} skupiliśmy się na badaniu możliwości zastosowania FL w wersji scentralizowanej do algorytmów SS. Jednak takie podejście wiąże się z~potrzebą scentralizowanej koordynacji procesu wykrywania widma. Jednak nie zawsze okazuje się to możliwe. W tym artykule przedstawiamy wyniki eksperymentu sprzętowego, w którym rozproszone uczenie maszynowe oparte na FL zostało zastosowane do ulepszenia algorytmów SS. W szczególności skupiliśmy się na sytuacji, w której jeden z węzłów całego systemu nie ma dostępu do danych treningowych.

\section{Definicja problemu}
Celem przeprowadzanego eksperymentu sprzętowego była praktyczna weryfikacja wydajności rozproszonego algorytmu detekcji obecności sygnału SS, w którym węzły współdzielą ze sobą parametry modeli AI, aby poprawić prawdopodobieństwo wykrycia zajętości widma. W szczególności rozważono przypadek, gdy pojedynczy węzeł jest wrażliwy na błędy detekcji ponieważ znajduje się w obszarze mierzonych niskich wartości stosunku mocy sygnału do mocy szumu (ang. Signal to Noise Ratio, SNR). Zgodnie z koncepcją rozproszonego uczenia federacyjnego, poprawa wydajności wykrywania obecności sygnału jest możliwa właśnie poprzez współdzielenie wyuczonych modeli AI, prowadząc do zwiększenia informacji o wykorzystaniu widma dzięki wiedzy od węzłów sąsiadujących. W przeprowadzonym eksperymencie laboratoryjnym każdy węzeł przeprowadza detekcję sygnału szerokopasmowego telewizji cyfrowej DVB-T, stosując w tym celu algorytm oparty na wartościach własnych.

\subsection{Gromadzenie danych}
\begin{figure}[H]
    \centering
    \includegraphics[width=0.48\textwidth]{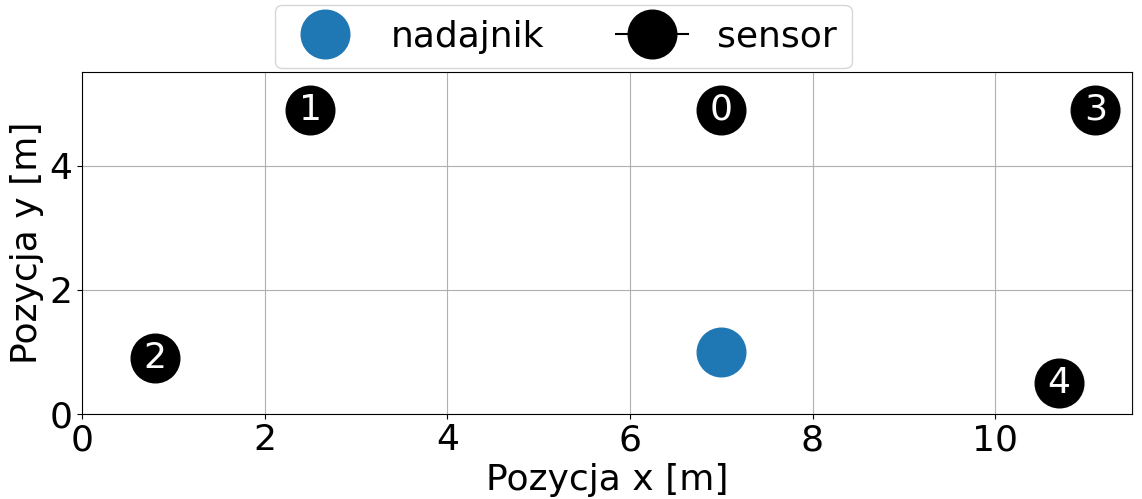}
    \caption{Topologia sieci używana podczas pomiarów.}
    \label{fig:topo}
\end{figure}

Próbki sygnału oraz szumu wykorzystane w tej pracy zostały zebrane w laboratorium, gdzie rozmieszczono pojedynczy nadajnik i pięć czujników tak jak to pokazano na Rys.~\ref{fig:topo}. 

Rolę każdego czujnika pełnił komputer stacjonarny z~oprogramowaniem GNU Radio zainstalowanym w~systemie operacyjnym Linux; jako stopień wejściowy odbiornika wykorzystano popularne układy radia programowalnego Universal Software Radio Peripheral (USRP) w wersji B210. Każdy odbiornik zbierał 10 milionów próbek IQ~w~trakcie pojedynczej sekwencji pomiarowej. W celu zapewnienia poprawnych danych do treningu modeli AI, każdy czujnik został uruchomiony 50 razy podczas transmisji sygnału DVB-T oraz 50 razy, gdy transmisja sygnału była wyłączona. Sygnał DVB-T był transmitowany przez generator sygnału Rohde\&Schwarz SMBV100A, dla założonej częstotliwości środkowej 2,0 GHz i mocy transmisyjnej zmienianej w zakresie od -40 dBm do 10 dBm z~krokiem o 1 dBm. Zebrane w ten sposób próbki zostały dalej wykorzystane w algorytmie detekcji zajętości widma.

\subsection{Przetwarzanie danych}

Przygotowanie próbek IQ zebranych przez czujniki zostało rozpoczęte od przefiltrowania pasma 8 MHz, w którym powinien być odbierany oczekiwany sygnał DVB-T. Następnie dla zbioru $L=8$ kolejnych wektorów próbek o~rozmiarze $N=32768$ obliczono średnią odbieraną moc $\mu$; wyznaczono także macierz korelacji $\mathbf{R}$, dla której wyznaczono wartości własne $\gamma_i$, gdzie $i=0,1,...L-1$, a~$L$ to liczba wartości własnych. Jako metrykę decyzyjną przyjęto stosunek maksymalnej do minimalnej wartości własnej $T=\frac{\gamma_{\max}}{\gamma_{\min}}$ oraz wartość autokorelacji. W rezultacie otrzymano zestaw wartości wykorzystywanych dalej jako dane wejściowe w procesie uczenia maszynowego. Dodatkowo dane te musiały zostać znormalizowane, dlatego w przypadku autokorelacji przeprowadzono normalizację min-max, a w przypadku wartości własnych i metryki decyzyjnej przeprowadzono normalizację z-score. Przetestowano wiele podejść i kombinacji, aby wybrać odpowiednie metody normalizacji dla poszczególnych danych w celu wygenerowania najlepszych wyników.

\section{Plan eksperymentu}

Podczas eksperymentu czujniki podejmują próbę detekcji obecności sygnału DVB-T. W scenariuszu referencyjnym każdy czujnik wykorzystuje jedynie próbki zebrane przez siebie w procesie tworzenia modelu uczenia maszynowego. Założono, że każdy czujnik tworzy własny model sieci neuronowej z dwiema warstwami (cztery węzły w każdej warstwie), a proces jest powtarzany dziesięć razy z różnymi podziałami danych treningowych i testowych (przy użyciu podejścia Stratified K-Fold). Przeprowadzone zostały również czasochłonne analizy, typowe dla algorytmów uczenia maszynowego, których celem było określenie liczby warstw i węzłów. Dwa główne wnioski z takich analiz to: brak zauważalnej poprawy dokładności modeli przy dodaniu trzeciej warstwy do modelu; zastosowanie większej liczby węzłów (około 20) w każdej z warstw nieznacznie poprawia dokładność modeli. Z uwagi na nieznaczną poprawę dokładności (maksymalnie kilka procent) oraz fakt wykorzystania uczenia federacyjnego, co wiąże się z~koniecznością wymiany współczynników modeli, dobrana została celowo względnie mała liczba węzłów, która jednak zapewnia dość dobre wyniki dokładności modeli. \\
Zwracając uwagę na główne wyzwanie w procesie detekcji zajętości widma, którym jest dostęp do danych treningowych, symulowana była sytuacja, w której jeden z czujników nie jest w stanie trenować modelu (nie ma dostępu do danych treningowych). W takiej sytuacji czujnik nieposiadający dostępu do danych treningowych wykorzystuje model utworzony z domyślnymi współczynnikami i~nie zmienia ich podczas symulacji.

\subsection{Uczenie federacyjne}
Aby umożliwić czujnikowi nieposiadającemu dostępu do danych treningowych adaptację modelu, rozważone zostało federacyjne podejście do uczenia, w którym współczynniki modelu AI są wymieniane z sąsiednimi czujnikami. W~każdej iteracji procesu uczenia federacyjnego obliczane są wagi współczynników sąsiednich czujników, aby określić, który z czujników może mieć lepszy model. W tym przypadku wzięte zostało uwagę kilka aspektów:
\begin{itemize}
    \item wpływ czasu uczenia - czujnik zaczyna z wagą 0,8 dla własnych współczynników i zwiększa tę wagę do prawie 1,0 pod koniec procesu uczenia; 
    \item wynik dokładności ostatniego kroku uczenia - waga własnych współczynników jest skalowana liniowo z ostatnim wynikiem dokładności, aby nadać priorytet etapom uczenia, gdy osiągnięty zostanie wysoki wynik dokładności własnego modelu lub nadać priorytet sąsiednim modelom, gdy osiągnięty zostanie niski wynik dokładności własnego modelu;
    \item w przypadku uwzględnienia wielu sąsiednich modeli ich wagi są obliczane przy użyciu ważenia odwrotnej odległości (ang. Inverse Distance Weighting , IDW).
\end{itemize}
W pracy rozważono różne współczynniki algorytmu ważenia odwrotnej odległości (dokładnie: 0, 1, 2, 3).

\section{Wyniki eksperymentu}
Uzyskane wyniki przedstawiono z dwóch perspektyw. Pierwsza dotyczy oceny jakości procesu uczenia maszynowego, a dokładniej oceny finalnego, wytrenowanego modelu w każdym z czujników. Podstawową metryką do stosowaną do tej oceny jest miara dokładności (ang. accuracy score), jednakże określona ona tylko liczbę poprawnych predykcji w stosunku do wszystkich prób. Najczęściej nie jest to wystarczająca miara i tak w przypadku detekcji zajętości widma szczególnie interesujące są błędy predykcji, gdyż fałszywy alarm nie powinien być traktowy na równi z sytuacją nie wykrycia obecnego sygnału. Z tego powodu do oceny finalnego modelu wykorzystana została metryka F1 (ang. F1 score), która z jednakową wagą uwzględnia w ocenie precyzję (ang. precision) oraz czułość (ang. recall).

\begin{figure}[H]
    \centering
    \includegraphics[width=0.48\textwidth]{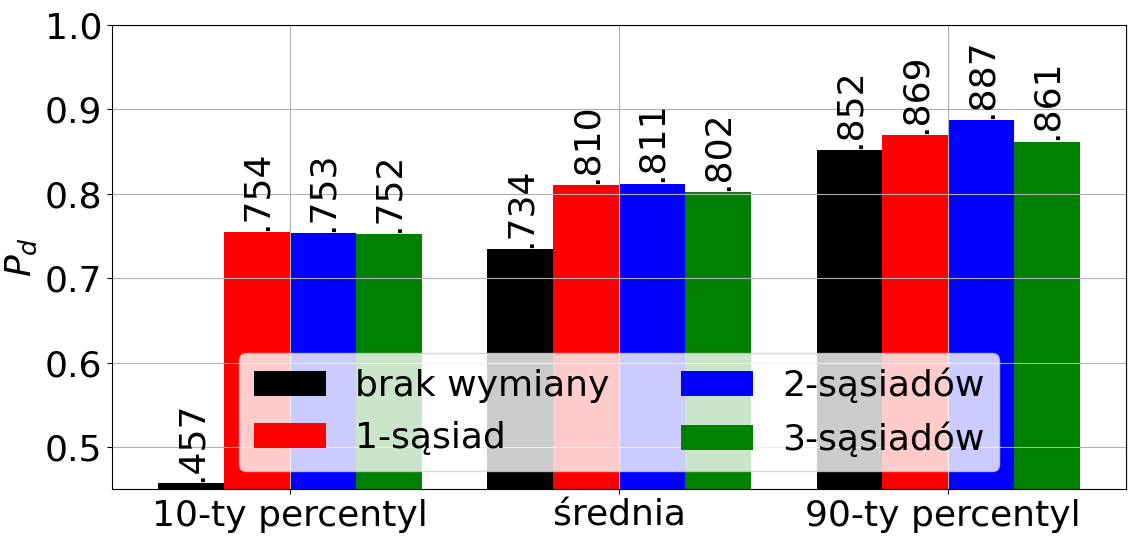}
    \caption{Prawdopodobieństwo detekcji (współczynnik IDW=0)}
    \label{fig:p_d_0}
\end{figure}
\begin{figure}[H]
    \centering
    \includegraphics[width=0.48\textwidth]{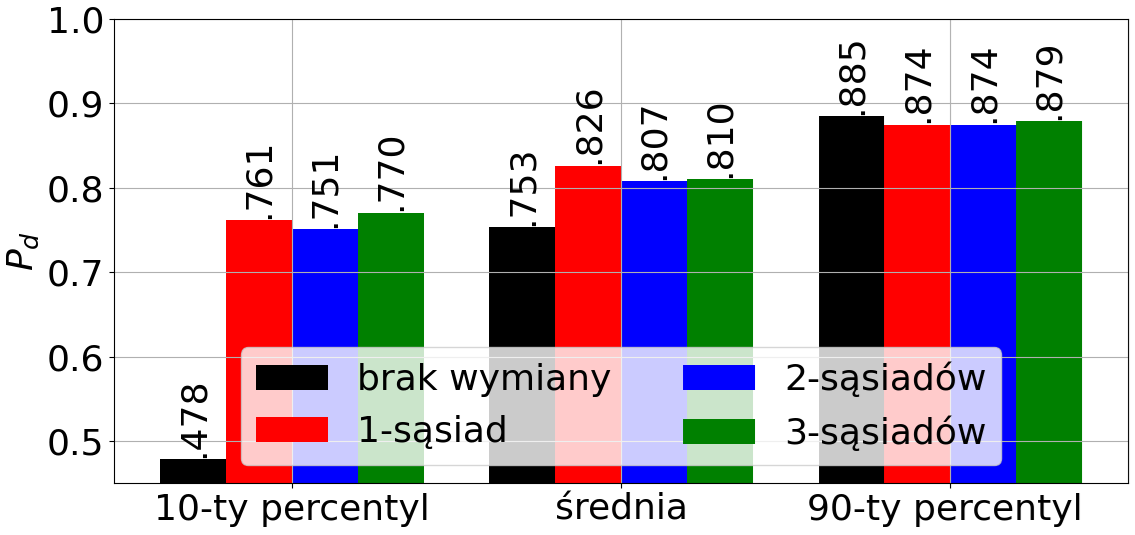}
    \caption{Prawdopodobieństwo detekcji (współczynnik IDW=1)}
    \label{fig:p_d_1}
\end{figure}
\begin{figure}[H]
    \centering
    \includegraphics[width=0.48\textwidth]{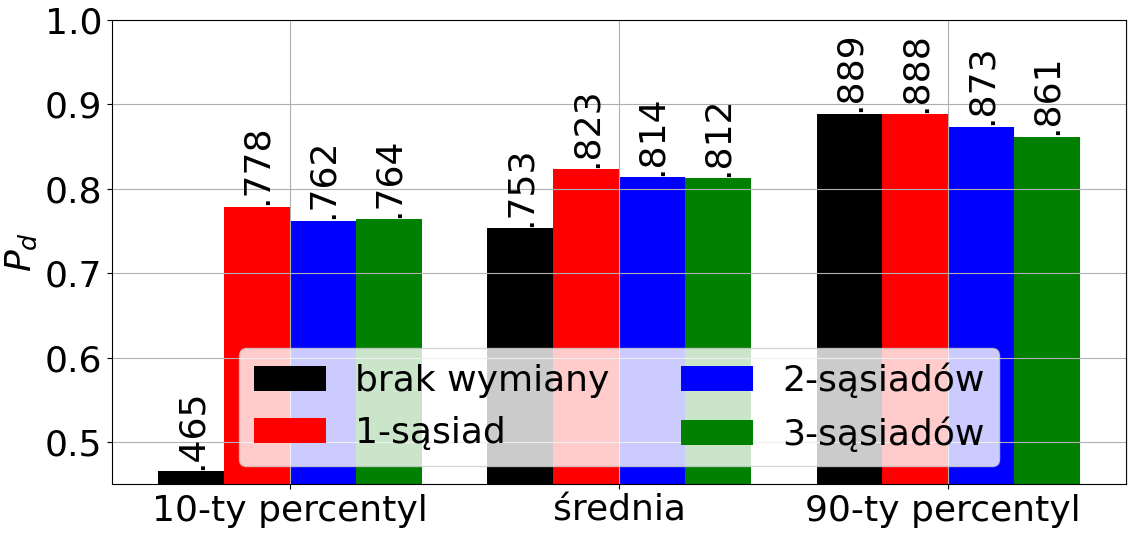}
    \caption{Prawdopodobieństwo detekcji (współczynnik IDW=2)}
    \label{fig:p_d_2}
\end{figure}
\begin{figure}[H]
    \centering
    \includegraphics[width=0.48\textwidth]{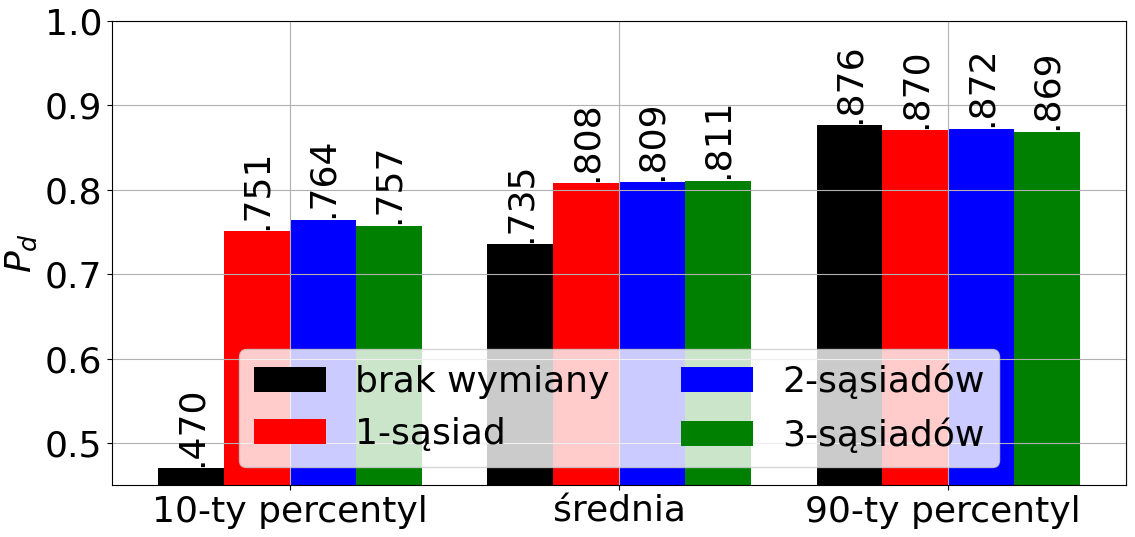}
    \caption{Prawdopodobieństwo detekcji (współczynnik IDW=3)}
    \label{fig:p_d_3}
\end{figure}

Precyzja modelu określa jaka część próbek z sygnałem udało się wykryć prawidłowo w drodze predykcji (liczba prawidłowych detekcji podzielona przez liczbę prawidłowych detekcji powiększoną przez liczbę nieprawidłowych detekcji), a czułość określa jaką część próbek sygnału pominięto w drodze predykcji (liczba prawidłowych detekcji podzielona przez liczbę prawidłowych detekcji powiększoną przez liczbę nieprawidłowych braków detekcji). 

Druga przyjęta perspektywa dotyczy oceny skuteczności detekcji zajętości widma. W tym przypadku pod uwagę wzięto prawdopodobieństwo detekcji sygnału przez czujnik. Z uwagi na strukturę symulacji, gdy podczas każdej iteracji inny czujnik nie ma dostępu do danych treningowych, wyniki przedstawiono w postaci średniego wyniku dla czujnika oraz 10-tego i 90-tego percentyla, co obrazuje jeden z najsłabszych wyników i jeden z najlepszych wyników (celowo pomijając minimum i maksimum).

Uzyskane prawdopodobieństwo detekcji pokazane zostało na Rys.~\ref{fig:p_d_0}~-~\ref{fig:p_d_3}. Warto zwrócić uwagę, że w analizowanym scenariuszu, gdy wszystkie czujniki znajdują się w tym samym pomieszczeniu, a warunki propagacyjne między nadajnikiem a każdym z czujników bardzo zbliżone to wymiana współczynników modeli między różną liczbą sąsiadów przynosi bardzo zbliżone efekty. Konsekwencją tego jest również mały wpływ współczynnika IDW na wyniki. Najważniejszym jednak z uzyskanych rezultatów wnioskiem jest zwiększenie prawdopodobieństwa detekcji najsłabszych czujników (niemających dostępu do danych treningowych) z przedziału 0.457-0.478 do przedziału 0.751-0.778 (w zależności od współczynnika IDW oraz liczby sąsiadów rozważanych w procesie wymiany współczynników modeli) - co w praktycznym zastosowaniu wyraźnie zwiększa szanse na dokonanie poprawnej detekcji obecności sygnału.

\begin{figure}[H]
    \centering
    \includegraphics[width=0.48\textwidth]{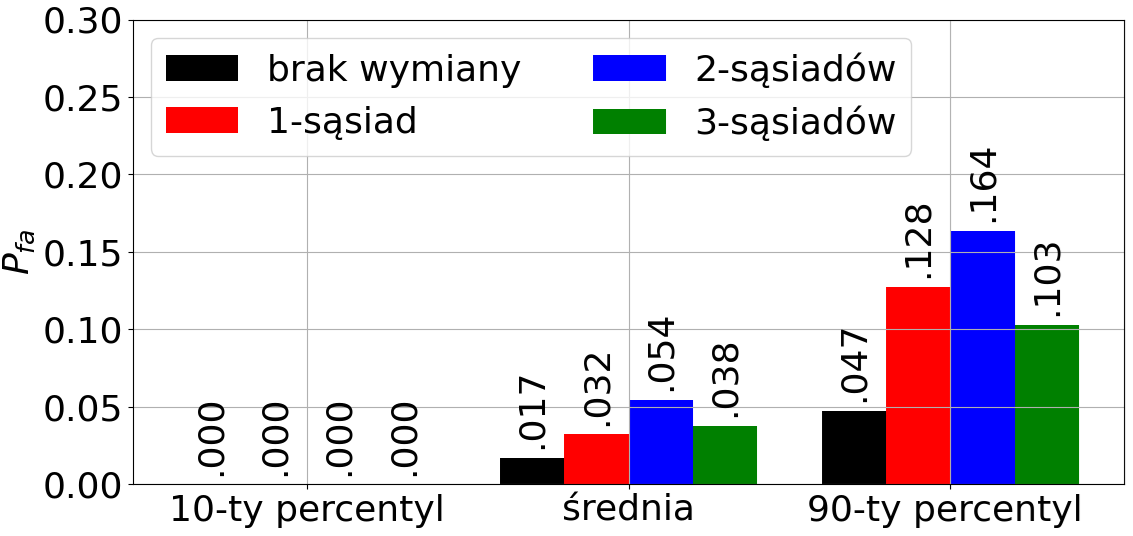}
    \caption{Prawdopodobieństwo fałszywego alarmu (współczynnik IDW=0)}
    \label{fig:p_fa_0}
\end{figure}

Jednocześnie warto przeanalizować wyniki prawdopodobieństwa fałszywego alarmu przedstawionego na Rys.~\ref{fig:p_fa_0} (tylko dla współczynnika IDW równego 0). Najgorsze z przypadków (tym razem reprezentowane w 90-tym percentylu wyników) zwiększają prawdopodobieństwo fałszywego alarmu z 0,047 do przedziału od 0,103 do 0,164. Warto pamiętać, że jest to wynik uzyskany jednocześnie z zwiększeniem prawdopodobieństwa detekcji z około 0,45 na około 0,75.

\begin{figure}[H]
    \centering
    \includegraphics[width=0.48\textwidth]{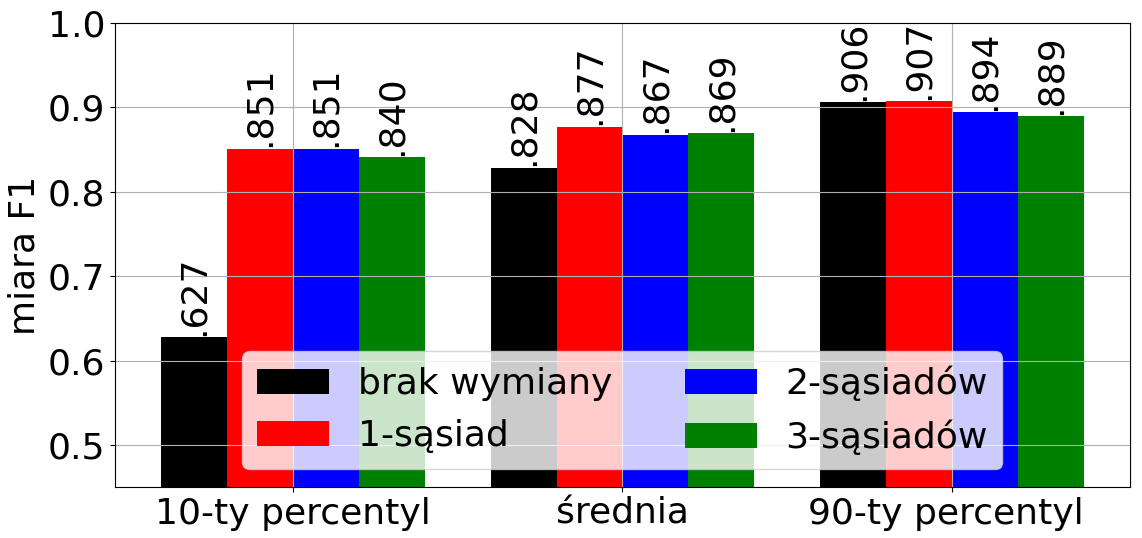}
    \caption{Miara F1 (współczynnik IDW=0)}
    \label{fig:f1_0}
\end{figure}

\begin{figure}[H]
    \centering
    \includegraphics[width=0.48\textwidth]{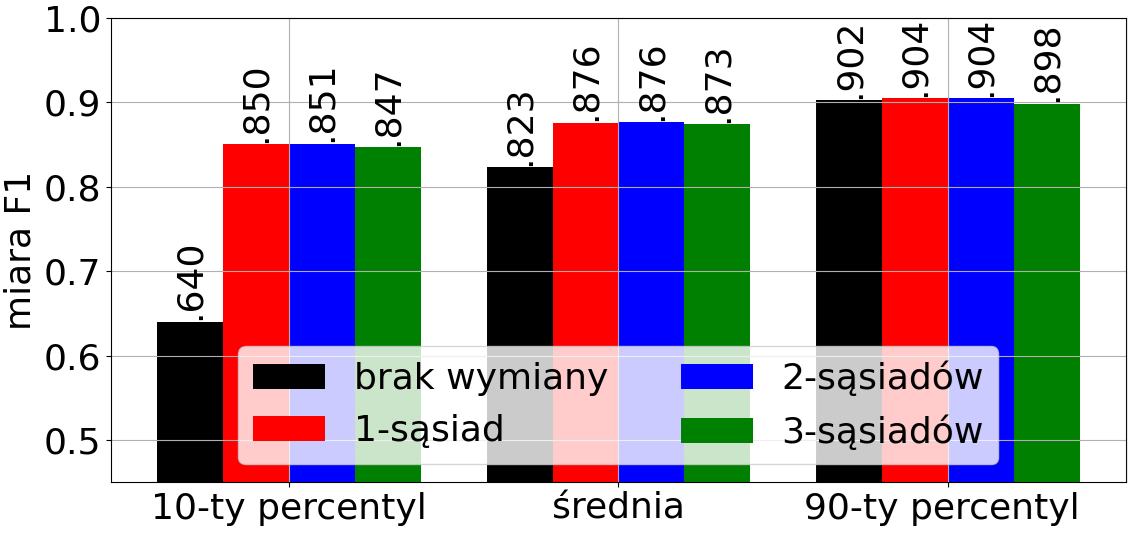}
    \caption{Miara F1 (współczynnik IDW=3)}
    \label{fig:f1_3}
\end{figure}

Wyniki jakości finalnych modeli pokazane zostały na Rys.~\ref{fig:f1_0}~-~\ref{fig:f1_3} w postaci uzyskanej metryki F1 (dla skrajnych wartości współczynnika IDW). Warto zauważyć, że czujniki, które nie mają możliwości podjęcia poprawnej decyzji (prawdopodobieństwo detekcji poniżej 50\%) posiadają modele ocenione (miarą F1) na około 0,65. Przy ocenie średniego modelu widać zdecydowanie, że wymiana współczynników z tylko jednym sąsiadem nie działa korzystnie i znacznie lepsze rezultaty można uzyskać poprzez wymianę z większą liczbą sąsiadów. Podobny trend odzwierciedlony jest w wynikach przedstawiających prawdopodobieństwo detekcji. Ponownie wpływ współczynnika IDW wykorzystywanego podczas tej symulacji jest pomijalny ze względu na umieszczenie wszystkich czujników w jednym pomieszczeniu. Jednakże w przypadku bardziej zróżnicowanego środowiska propagacyjnego odpowiednie kryterium doboru sąsiadów a następnie wag ich modelu byłoby kluczowe do poprawnego działania przedstawianego rozwiązania.

% \begin{figure}[H]
%     \centering
%     \includegraphics[width=0.48\textwidth]{f1_1}
%     \caption{Miara F1 (współczynnik IDW=1)}
%     \label{fig:f1_1}
% \end{figure}

% \begin{figure}[H]
%     \centering
%     \includegraphics[width=0.48\textwidth]{f1_2}
%     \caption{Miara F1 (współczynnik IDW=2)}
%     \label{fig:f1_2}
% \end{figure}

\begin{figure}[H]
    \centering
    \includegraphics[width=0.48\textwidth]{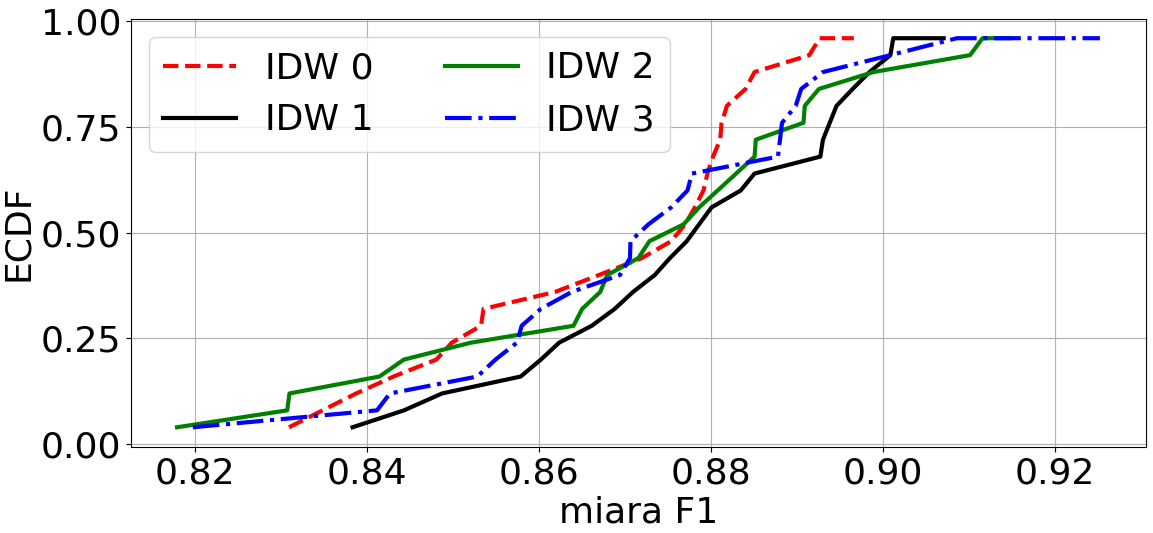}
    \caption{Dystrybuanta miary F1}
    \label{fig:idw_f1}
\end{figure}

Wpływ współczynnika IDW na miarę F1 w scenariuszu wymiany modeli między trzema sąsiadami przedstawiono dodatkowo na Rys.~\ref{fig:idw_f1} w formie dystrybuanty. Zauważalnym jest mniejsza rozbieżność miary F1 (między około 0,83 a 0,89) gdy każdy sąsiad ma taką samą wagę (IDW 0), a największą rozbieżność (między około 0,82 a~0,93) gdy zastosowano IDW z współczynnikiem 3. Zastosowanie większych współczynników IDW daje zatem możliwość lepszego dostosowania modelu do zastanej sytuacji, jednak wraz z zwiększonym ryzykiem uzyskania gorszego modelu. Zastosowanie mniejszych współczynników IDW powoduje większe uśrednianie modelu co minimalizuje ryzyko uzyskania słabszego modelu, ale również nie pozwala na osiągnięcie lepszych wyników lokalnie.

\section{Podsumowanie}
W pracy przedstawiono wyniki przeprowadzonego eksperymentu sprzętowego, w którym zastosowano rozproszone uczenie federacyjne dla poprawy detekcji obecności sygnału w obserwowanym widmie częstotliwościowym. Udało się wykazać, że w wyniku uczenia federacyjnego, które opiera się na wymianie współczynników modelu uczenia maszynowego, proces uczenia może zostać usprawniony - szczególnie w przypadku czujników z utrudnionym dostępem do danych treningowych. W ramach dalszych prac planowane jest włączenie do analizy dodatkowego czujnika, którego praca został celowo zaburzona przez niedokładne podłączenie anteny odbiorczej, a także analiza wpływu liczby wymienianych próbek (liczba węzłów w każdej z warstw modelu) na skuteczność działania całego systemu oraz porównanie wyników z systemem wykrywania zajętości widma z węzłem centralnym.

\section*{Podziękowanie}
Praca powstała w ramach projektu finansowanego przez Narodowe Centrum Nauki (nr 2021/41/N/ST7/01298).

\end{multicols}
\end{document}